# Split Kinetic Energy Method for Quantum Systems with Competing Potentials

H. Mineo and Sheng D. Chao*

Institute of Applied Mechanics, National Taiwan University, Taipei 106, Taiwan

(May 6, 2012)

*sdchao@iam.ntu.edu.tw

Abstract

For quantum systems with competing potentials, the conventional perturbation theory often yields an asymptotic series and the subsequent numerical outcome becomes uncertain. To tackle such kind of problems, we develop a general solution scheme based on a new energy dissection idea. Instead of dividing the potential energy into "unperturbed" and "perturbed" terms, a partition of the kinetic energy is performed. By distributing the kinetic energy term in part into each individual potential, the Hamiltonian can be expressed as the sum of the subsystem Hamiltonians with respective competing potentials. The total wavefunction is expanded by using a linear combination of the basis sets of respective subsystem Hamiltonians. We first illustrate the solution procedure using a simple system consisting of a particle under the action of double $\delta$-function potentials. Next, this method is applied to the prototype systems of a charged harmonic oscillator in strong magnetic field and the hydrogen molecule ion. Compared with the usual perturbation approach, this new scheme converges much faster to the exact solutions for both eigenvalues and eigenfunctions. When properly extended, this new solution scheme can be very useful for dealing with strongly coupling quantum systems.

*I. Introduction*

Basis set expansion methods have been a general solution scheme to the quantum eigenvalue problems ever since the dawn of quantum mechanics [1]. In the perturbation theory the system Hamiltonian is often divided into "unperturbed" and "perturbed" terms [2-4]. The success of a perturbation series expansion relies on the properly chosen "unperturbed" Hamiltonian (e.g. they are exactly solved or well-represented analytically) and smallness of the "perturbed" potential. It is thus required to determine the dominant potential term in the full Hamiltonian. However, there exist many situations where the individual potentials are equally important, or competing, in "strong-coupling" regions. As of this, an initially thought "weak" interaction would become dominant and the series converges slowly or even diverges. Although there exist re-summation methods such as large-order perturbation theory [5] to mitigate this difficulty, the solution scheme becomes very tedious and the numerical outcomes are often uncertain. One such example is an atom in strong magnetic fields [6]. It is well known that when the magnetic field strength is comparable to the atomic Coulomb potential, the usual perturbation series becomes asymptotic. In fact the system changes magnetic characteristics from the spherical symmetry (diamagnetic) to the cylindrical symmetry (paramagnetic). Therefore, one needs to go beyond the single perturbation theory to solve the problem.

There are several alternative methods to remedy the above problem. For example, the double perturbation theory uses two perturbation parameters to gauge the relative magnitudes of the competing potentials [7]. This theory works best when the two potentials can be simultaneously assigned to be "weak". In the alternating perturbation theory, the perturbation series is formed by utilizing the two interaction potentials alternately [8], yielding the so-called strong field approximation. The mixing-mode perturbation theory doubles the Hilbert space by using both basis sets of the two limiting Hamiltonians [9], which are equally "strong". All these above approaches rely on the proper division of the potential energy part *a priori*. However, because of the inhomogeneous distribution of the potentials, a "weak" interaction in one region might become a "strong" one in another region. Therefore, these split potential methods, albeit improves on, are still plagued with the original problem to some extent.

In this paper a new solution scheme is developed by considering the possibility of splitting the kinetic energy. At first glance one might wonder how the kinetic energy, which is a differential operator anyway, can be separated. In fact, we do not split the operator itself. Instead, we adjust the mass parameter. We start by attributing an effective mass term into each partial kinetic energy such that the total kinetic energy remains unchanged. Next, each partitioned kinetic energy term combines with an individual potential function to form a subsystem. The system Hamiltonian is thus written as the sum of subsystem Hamiltonians. This approach is most useful if the problem is solvable for each subsystem but not for the whole system. In this case, the subsystem solutions can thus be used by simply replacing the original mass parameters by the effective ones. One apparent advantage of this approach is that it is not necessary to determine the relative importance of each potential energy term *a priori*. In this sense, the kinetic energy has been equally distributed to each potential energy term. Therefore, the solution scheme is more "democratic" and suitable for "strong-coupling" systems. Because it is the kinetic energy to be divided, we will call this approach the kinetic energy partition (KEP) method and the results using this method the KEP solutions.

The organization of this paper is as follows. In Sec. II we derive the basic equations based on the KEP scheme and discuss its main features. In Sec. III we illustrate the solution procedure using a simple model consisting of two $\delta$-function potentials. In Sec. IV we apply the KEP method to solve the Schroedinger equation for a charged harmonic oscillator in strong magnetic field. In Sec. V, the prototype system of the hydrogen molecule ion ($H_2^+$) is studied. We summarize this work and provide future perspectives in Sec. VI. In Appendix I we develop a perturbative approach based on the KEP scheme and in Appendix II we formulate the time dependent KEP method.

*II. Theoretical formulation of the KEP method*

In this section we derive the working equations based on the KEP idea. Without losing generality, we consider a system of one particle of mass *m* with two competing potentials. The Hamiltonian of the system is

$$\hat{H} = \hat{T} + V_1 + V_2 \tag{1}$$

where $\hat{T} = \frac{\hat{p}^2}{2m}$ is the kinetic energy, and $V_1$ and $V_2$ are the two competing potentials. We divide the kinetic energy into two terms by assigning an effective mass which is twice the original mass.

$$\hat{T} = \hat{T}_1 + \hat{T}_2 \tag{2a}$$

$$\hat{T}_1 = \frac{\hat{p}^2}{2(2m)} \tag{2b}$$

$$\hat{T}_2 = \frac{\hat{p}^2}{2(2m)} \tag{2c}$$

With this partition, the total Hamiltonian is written as

$$\hat{H} = \hat{H}_1 + \hat{H}_2 \tag{3a}$$

where the subsystem Hamiltonians are

$$\hat{H}_1 = T_1 + V_1 = \frac{\hat{p}^2}{2(2m)} + V_1 \tag{3b}$$

$$\hat{H}_2 = T_2 + V_2 = \frac{\hat{p}^2}{2(2m)} + V_2 \tag{3c}$$

Assume the eigenvalue problems for the two subsystems have been obtained

$$\hat{H}_1\psi_{1n} = E_{1n}\psi_{1n} \tag{4a}$$

$$\hat{H}_2\psi_{2k} = E_{2k}\psi_{2k} \tag{4b}$$

where $n$ =1, 2, 3,..., $N$ and $k$ =1, 2, 3, ..., $K$ are the respective quantum numbers. We expand the wavefunction using a linear combination of both basis sets

$$\psi = \sum_n C_{1n}\psi_{1n} + \sum_k C_{2k}\psi_{2k} \tag{5}$$

where $C_{1n}$ and $C_{2k}$ are the expansion coefficients to be determined. Substituting this series into the Schroedinger equation

$$\hat{H}\psi = E\psi \tag{6}$$

multiplying $C_{1m}^{*}$ and $C_{2l}^{*}$ respectively and integrating over the coordinate space, we have,

$$C_{1m}(E_{1m} - E) + \sum_n C_{1n} \langle \psi_{1m} | \hat{H}_2 | \psi_{1n} \rangle + \sum_k C_{2k}(E_{2k} + E_{1m} - E)\langle \psi_{1m} | \psi_{2k} \rangle = 0 \tag{7a}$$

$$C_{2\ell}(E_{2\ell} - E) + \sum_k C_{2k} \langle \psi_{2\ell} | \hat{H}_1 | \psi_{2k} \rangle + \sum_n C_{1n}(E_{1n} + E_{2\ell} - E)\langle \psi_{2\ell} | \psi_{1n} \rangle = 0 \tag{7b}$$

Notice that

$$\hat{H}_2 = \hat{T}_2 + V_2 = \hat{T}_1 + V_1 + \left(V_2 - V_1\right)$$
$$\hat{H}_1 = \hat{T}_1 + V_1 = \hat{T}_2 + V_2 + \left(V_1 - V_2\right)$$

We thus can rewrite the coupled equations as

$$C_{1m}(2E_{1m} - E) + \sum_n C_{1n} \left\langle \psi_{1m} \middle| (V_2 - V_1) \middle| \psi_{1n} \right\rangle + \sum_k C_{2k} (E_{2k} + E_{1m} - E) \left\langle \psi_{1m} \middle| \psi_{2k} \right\rangle = 0 \tag{8a}$$

$$C_{2\ell}(2E_{2\ell} - E) + \sum_k C_{2k} \left\langle \psi_{2\ell} \middle| (V_1 - V_2) \middle| \psi_{2k} \right\rangle + \sum_n C_{1n} (E_{1n} + E_{2\ell} - E) \left\langle \psi_{2\ell} \middle| \psi_{1n} \right\rangle = 0 \tag{8b}$$

Notice that the coupling interaction terms depend on the difference of the potentials, which is the main feature of using the KEP method. Thanks to this, if the two potentials are equally large in magnitude, the two subsystems are formally "separated". This is best illustrated by the following trivial example. Let's consider a particle under the action of two identical potentials.

$$V_1 = V_2$$

In this case, the coupled equations becomes

$$C_{1m}(2E_{1m} - E) + C_{2m}(E_{2m} + E_{1m} - E) = 0 \tag{9a}$$

$$C_{2\ell}(2E_{2\ell} - E) + C_{1\ell}(E_{1\ell} + E_{2\ell} - E) = 0 \tag{9b}$$

Therefore the KEP solutions are

$$E_m^{KEP} = E_{1m} + E_{2m} \tag{10}$$

which are the exact solutions. For $V_1(x) \approx V_2(x)$ in some spatial regions, we can develop a perturbative approach, which will be discussed in Appendix I. Likewise, the time-dependent version of the KEP method can be formulated in a similar way as shown in Appendix II.

*III. A simple model: double $\delta$-function potentials*

To illustrate the solution procedure, in this section we use a simple model which consists of two $\delta$-function potentials with the same well depth $\lambda$ at positions $x = \pm a$. For the single $\delta$-function potential cases, the exact solutions are known

$$V_1(x) = -\lambda\delta(x-a) \tag{11}$$

$$\psi_1 = \begin{cases} \sqrt{k}e^{k(x-a)} & x < a \\ \sqrt{k}e^{-k(x-a)} & x > a \end{cases} \tag{12}$$

and

$$V_2(x) = -\lambda\delta(x+a) \tag{13}$$

$$\psi_2 = \begin{cases} \sqrt{k}e^{k(x+a)} & x < -a \\ \sqrt{k}e^{-k(x+a)} & x > -a \end{cases} \tag{14}$$

$$E_1 = -\frac{m^*\lambda^2}{2\hbar^2} = E_2 \tag{15}$$

Here $k = \frac{m^*}{\hbar^2}\lambda = \sqrt{\frac{-2m^*E_1}{\hbar^2}}$ and $m^* = 2m$. Also for the double $\delta$-function potential, we have

$$V = V_1 + V_2,$$

$$\psi = \begin{cases} \dfrac{\sqrt{2K}\cosh\gamma}{\sqrt{e^{2\gamma}+2\gamma+1}} e^{-K(|x|-a)} & |x| > a \\ \dfrac{\sqrt{2K}}{\sqrt{e^{2\gamma}+2\gamma+1}} \cosh Kx & |x| < a \end{cases} \tag{16}$$

$$E = -\frac{\hbar^2}{2m}K^2 \tag{17}$$

where $\beta \equiv 2a \cdot \frac{m}{\hbar^2}\lambda$ and $\gamma \equiv Ka$. The equation $\gamma(1+\tanh\gamma) = \beta$ is used to determine $K$ or equivalently the energy.

Using the KEP method, let`s consider a two-state approximation

$$C_1(2E_1 - E) + C_1\langle\psi_1|(V_2 - V_1)|\psi_1\rangle + C_2(E_2 + E_1 - E)\langle\psi_1|\psi_2\rangle = 0 \qquad (18a)$$

$$C_2(2E_2 - E) + C_2\langle\psi_2|(V_1 - V_2)|\psi_2\rangle + C_1(E_1 + E_2 - E)\langle\psi_2|\psi_1\rangle = 0 \qquad (18b)$$

The necessary matrix elements can all be calculated and represented analytically

$$\xi \equiv \langle\psi_2|(V_1 - V_2)|\psi_2\rangle = \lambda k(1 - e^{-4ka}) = \langle\psi_1|(V_2 - V_1)|\psi_1\rangle \qquad (19)$$

and

$$\eta \equiv \langle\psi_1|\psi_2\rangle = e^{-2ka}(2ka + 1) \qquad (20)$$

We then obtain

$$\begin{vmatrix} 2E_1 + \xi - E & \eta(2E_1 - E) \\ \eta(2E_1 - E) & 2E_1 + \xi - E \end{vmatrix} = 0 \qquad (21)$$

or

$$E = 2E_1 + \frac{\xi}{1 \pm \eta} \equiv -\frac{\hbar^2}{2m}K_{KEP}^2 \qquad (22)$$

or

$$K^{KEP} = \sqrt{\frac{2m}{\hbar^2}\cdot\frac{m^*}{\hbar^2}\lambda^2\left(1 - \frac{1 - e^{-4ka}}{1 + e^{-2ka}(2ka + 1)}\right)} \qquad (23)$$

Here the minus-sign solution is excluded because it yields an unbound state. We compare the $K^{KEP}$ with the exact $K$ in Fig. 1. As can be seen, even using only two states the KEP energy has been close to the exact solution, much so for smaller $\lambda$. In Fig. 2 we compare the KEP wavefunction with the exact solution. Also we see very good agreement.

*IV. The KEP method applied to a harmonic oscillator in strong magnetic field*

In this section we apply the KEP method to a charged particle with the reduced mass $\mu$ and the charge $e$ under the action of a harmonic potential in a strong magnetic field. Using the polar coordinates ($\rho$, $\varphi$) with its origin at the equilibrium of oscillation in the magnetic field $H$ pointing to the z direction, the Hamiltonian has the form,

$$\hat{H} = -\frac{1}{2\mu}\left\{\frac{1}{\rho}\frac{\partial}{\partial\rho}\left(\rho\frac{\partial}{\partial\rho}\right)+\frac{1}{\rho^2}\frac{\partial^2}{\partial^2\varphi}\right\}+\frac{ieH}{2c\mu}\frac{\partial}{\partial\varphi}+V_1(\rho)+V_2(\rho) \tag{24}$$

where $V_1(\rho)=\frac{e^2H^2\rho^2}{8c^2\mu}$, $V_2(\rho)=\frac{\mu\omega^2\rho^2}{2}$, and $\omega$ is the frequency of the harmonic oscillator. The total wavefunction is given by

$$\Psi(\rho,\varphi)=\sqrt{N_c}\exp(im\varphi)f(\rho) \qquad (m=0,\ \pm1,\ \pm2,\ \cdots) \tag{25}$$

The function $f(\rho)$ satisfies the following equation,

$$f''+\frac{1}{\rho}f'+\left[2\mu E-m\frac{H}{c}-\frac{m^2}{\rho^2}-\frac{H^2\rho^2}{4c^2}-\mu^2\omega^2\rho^2\right]f=0 \tag{26}$$

Using the characteristic lengths $a=\sqrt{\frac{c}{H}}$, $b=\left(\frac{H^2}{4c^2}+\mu^2\omega^2\right)^{-1/4}$, and the variable $\xi=\rho^2/b^2$, we have,

$$f''(\xi)+\frac{1}{\xi}f'(\xi)+\frac{1}{4}\left(\frac{2\mu Eb^2}{\xi}-m\frac{b^2}{a^2\xi}-\frac{m^2}{\xi^2}-1\right)f(\xi)=0. \tag{27}$$

The solution of this equation is given as

$$f(\xi)=\exp(-\xi/2)\xi^{|m|/2}M(-n,|m|+1,\xi) \tag{28}$$

with the definition of the quantum number

$$-n = \frac{|m|+1}{2} + \frac{m}{4a}b^2 - \frac{2\mu E}{4}b^2 \qquad (n = 0,\ 1, 2,\ \cdots) \tag{29}$$

where $M$ is the confluent hypergeometric function of the first kind. Thus the total wavefunction yields

$$\Psi_{n,m}(\rho,\varphi) = \sqrt{N_c}\,\exp\left(im\varphi - \frac{\rho^2}{2b^2}\right)\left(\frac{\rho}{b}\right)^{|m|} M\left(-n, |m|+1, \frac{\rho^2}{b^2}\right) \tag{30}$$

with $N_c = \dfrac{\left(n+|m|\right)!}{\pi b^2 \left(|m|!\right)^2 n!}$, and the energy is

$$E = (2n + |m| + 1)\sqrt{\omega^2 + {\omega_L}^2} + m\omega_L, \tag{31}$$

where $\omega_L = \dfrac{H}{2\mu c}$ is the angular velocity of the Larmor precession. For the case $m$=0, the Hamiltonian reduces to the form,

$$\hat{H} = \hat{T} + V_1(\rho) + V_2(\rho) \tag{32}$$

with $\hat{T} = -\dfrac{1}{2\mu}\left\{\dfrac{1}{\rho}\dfrac{\partial}{\partial\rho}\left(\rho\dfrac{\partial}{\partial\rho}\right)\right\}$ and the total wavefunction reduces to

$$\Psi_n(\rho) = \frac{1}{b\sqrt{\pi}}\exp\left(-\frac{\rho^2}{2b^2}\right)L_n\left(\frac{\rho^2}{b^2}\right) \tag{33}$$

where $L$ is the Laguerre function.

Using the KEP method, $\hat{H} = \hat{H}_1 + \hat{H}_2$ where

$$\hat{H}_1 = \frac{1}{2}\hat{T} + V_1(\rho) = -\frac{1}{2\mu^*}\left\{\frac{1}{\rho}\frac{\partial}{\partial\rho}\left(\rho\frac{\partial}{\partial\rho}\right)\right\} + \frac{e^2 H^{*2}\rho^2}{8c^2\mu^*} \tag{34}$$

and

$$\hat{H}_2 = \frac{1}{2}\hat{T} + V_2(\rho) = -\frac{1}{2\mu^*}\left\{\frac{1}{\rho}\frac{\partial}{\partial\rho}\left(\rho\frac{\partial}{\partial\rho}\right)\right\} + \frac{\mu^*\omega^{*2}\rho^2}{2} \quad (35)$$

with $\mu^* = 2\mu,\ H^* = \sqrt{2}H$ and $\omega^* = \omega/\sqrt{2}$. The subsystem energies of $\hat{H}_1$ and $\hat{H}_2$ are $E_{1n} = (2n+1)\omega_L^*$ and $E_{2n} = (2n+1)\omega^*$ with $\omega_L^* = \frac{H^*}{2\mu^* c} = \omega_L/\sqrt{2}$ respectively. The subsystem wavefunctions of $\hat{H}_1$ and $\hat{H}_2$ are also known

$$\Psi_{1n}(\rho) = \frac{1}{b_1\sqrt{\pi}}\exp\left(-\frac{\rho^2}{2b_1^{\ 2}}\right)L_n\left(\frac{\rho^2}{b_1^{\ 2}}\right) \quad (36a)$$

$$\Psi_{2n}(\rho) = \frac{1}{b_2\sqrt{\pi}}\exp\left(-\frac{\rho^2}{2b_2^{\ 2}}\right)L_n\left(\frac{\rho^2}{b_2^{\ 2}}\right) \quad (36b)$$

where $b_1 = \sqrt{\frac{2c}{H^*}}$ and $b_2 = \frac{1}{\sqrt{\mu^*\omega^*}}$. Using the dimensionless parameters $\mu = 1,\ \omega = \sqrt{2},\ H = 4c$ and $\omega_L = 2$, in Fig. 3 we compare the KEP energies for $N$=$K$=1~5 with the exact energies, together with those determined the respective subsystem Hamiltonians. It is clearly seen that by increasing $N$, the KEP energy levels converge to the exact solutions very fast. Normally we see that for $\mathrm{n} \le N-1$ the relative errors between the KEP and the exact energies are well below 5 %. Therefore, numerically the KEP energy levels are almost exact for the levels $\mathrm{n} \le N-1$. In Fig. 4 we present the total wavefunctions for the first four states (n=0 to n=3). The KEP wavefunctions are all in good agreement with the exact results for increasing $N$.

*V. The KEP method applied to the hydrogen molecule ion (*$H_2^+$)

In order to consider a more realistic case, we take the hydrogen molecule ion ($H_2^+$) as an example. The Hamiltonians $\hat{H}_i$ within the adiabatic approximation are given as

$$\hat{H}_i = \frac{\hat{p}^2}{2M_i} - \frac{1}{r_i}, \qquad (i = 1, 2) \tag{37}$$

with $M_1 = \alpha M$ and $M_2 = \frac{\alpha}{\alpha - 1} M$. Here $\alpha$ is a parameter which is introduced for a different mass-splitting way in the KEP theory. In previous sections we choose a constant value $\alpha = 2$ and each mass is the same; here $\alpha$ is a function of internuclear distance $R = |\vec{r}_1 - \vec{r}_2|$. $r_i = |\vec{r} - \vec{r}_i|$ is the distance from the nucleus *i*, and $\vec{r}_1 + \vec{r}_2 = \vec{0}$.

Here we consider the *s*-orbitals in the basis set, the eigenfunctions and eigenvalues of the Schroedinger equations $\hat{H}_i \psi_{in} = E_{in} \psi_{in}$ are given as

$$\psi_{in}(\vec{r}) = \frac{1}{\sqrt{4\pi}} R_{n0}(r_i / a_i), \tag{38}$$

$$E_{in} = -\frac{1}{2a_i n^2}, \tag{39}$$

where $n = 1, 2, \cdots$, $a_1 = 1/M_1 = \alpha^{-1}$ and $a_2 = 1/M_2 = 1 - \alpha^{-1}$. Then the coupled KEP equation (Eq. (8)) is modified as

$$C_{1m}(\alpha E_{1m} - E) + \sum_n C_{1n} \langle \psi_{1m} | (V_2 - (\alpha - 1) V_1) | \psi_{1n} \rangle + \sum_k C_{2k} (E_{2k} + E_{1m} - E) \langle \psi_{1m} | \psi_{2k} \rangle = 0 \tag{40a}$$

$$C_{2\ell}(\alpha E_{2\ell} - E) + \sum_k C_{2k} \langle \psi_{2\ell} | (V_1 - (\alpha - 1) V_2) | \psi_{2k} \rangle + \sum_n C_{1n} (E_{1n} + E_{2\ell} - E) \langle \psi_{2\ell} | \psi_{1n} \rangle = 0 \tag{40b}$$

The terms in Eq. (40) can be calculated by using the elliptical coordinates, $\xi = \frac{r_1 + r_2}{R}$ and $\eta = \frac{r_1 - r_2}{R}$, and we have,

$$\langle \psi_{1m} | (V_2 - (\alpha - 1)V_1) | \psi_{1n} \rangle$$

$$= \int_1^\infty d\xi \int_{-1}^1 d\eta \int_0^{2\pi} d\phi \frac{R^3}{8} \left(\xi^2 - \eta^2\right) \frac{1}{4\pi} R_{m0}\left(r_1 / a_1\right) R_{n0}\left(r_1 / a_1\right) \left( \frac{\alpha - 1}{r_1} - \frac{1}{r_2} \right)$$

$$= \int_1^\infty d\xi \int_{-1}^1 d\eta \frac{R^2}{8} \left(\xi^2 - \eta^2\right) R_{m0}\left( \frac{R(\xi + \eta)}{2a_1} \right) R_{n0}\left( \frac{R(\xi + \eta)}{2a_1} \right) \left( \frac{\alpha - 1}{\xi + \eta} - \frac{1}{\xi - \eta} \right), \qquad (41a)$$

$$\langle \psi_{2m} | (V_1 - V_2) | \psi_{2n} \rangle$$

$$= \int_1^\infty d\xi \int_{-1}^1 d\eta \int_0^{2\pi} d\phi \frac{R^3}{8} \left(\xi^2 - \eta^2\right) \frac{1}{4\pi} R_{m0}\left(r_2 / a_2\right) R_{n0}\left(r_2 / a_2\right) \left( \frac{\alpha - 1}{r_2} - \frac{1}{r_1} \right)$$

$$= \int_1^\infty d\xi \int_{-1}^1 d\eta \frac{R^2}{8} \left(\xi^2 - \eta^2\right) R_{m0}\left( \frac{R(\xi + \eta)}{2a_2} \right) R_{n0}\left( \frac{R(\xi + \eta)}{2a_2} \right) \left( \frac{1}{\xi + \eta} - \frac{\alpha - 1}{\xi - \eta} \right), \qquad (41b)$$

and

$$\langle \psi_{1m} | \psi_{2n} \rangle$$

$$= \int_1^\infty d\xi \int_{-1}^1 d\eta \frac{R^3}{16} \left(\xi^2 - \eta^2\right) R_{m0}\left( \frac{R(\xi + \eta)}{2a_1} \right) R_{n0}\left( \frac{R(\xi - \eta)}{2a_2} \right). \qquad (42a)$$

$$\langle \psi_{2m} | \psi_{1n} \rangle$$

$$= \int_1^\infty d\xi \int_{-1}^1 d\eta \frac{R^3}{16} \left(\xi^2 - \eta^2\right) R_{m0}\left( \frac{R(\xi + \eta)}{2a_2} \right) R_{n0}\left( \frac{R(\xi - \eta)}{2a_1} \right). \qquad (42b)$$

In Fig. 5 the ground state potential energy curves (energy as a function of internuclear distance *R)* calculated by Bowen et al. [10], Gaussian09 program package [11] using the HF/6-311++G(d,p) basis set, and the conventional linear combination of atomic orbitals as molecular orbitals (LCAO-MO) method [12] are shown, together with the KEP calculation with the 1s and 2s orbitals. We see that the KEP energy is consistent with both the analytically and numerically determined exact potential energies. On the other hand the conventional LCAO-MO method yields poor agreement with the exact solutions. Notice that all the calculated results approach to the -0.5 a.u. limit for large *R*. To provide a benchmark of the calculation in Table 1 we list a comparison of the equilibrium bond distances and bond energies calculated by several different methods [10-16] and the experimental data [17]. In Fig. 6 the $H_2^+$ ground state wavefunctions calculated by Bowen et al. [10], Gaussian09 program package [11], and the KEP method at the nuclear distance $R$=1.0 a.u. and $R$=2.0 a.u. where $y$=$z$=0, are plotted. We see an overall agreement between the KEP and the exact wavefunctions although the KEP results exhibit cusps at $x = \pm R/2$, and the absolute values are smaller at $x = 0$.

VI. Concluding Remarks

In this paper we develop a new energy dissection method for quantum eigenvalue problems with competing interaction potentials. Instead of the usual division of the competing potentials, we split the kinetic energy part such that the system Hamiltonians can be written as the sum of the subsystem Hamiltonians. We demonstrate the solution procedure using a simple system and then apply the method to a charged harmonic oscillator in strong magnetic fields and the hydrogen molecule ion. Both calculated energies and wavefunctions are approaching to the exact solutions using only a few number of basis functions. This new solution scheme, when properly extended, should be universally applicable to any quantum eigenvalue problem of such kind. The results presented in this paper, albeit illustrative, are believed to be very useful for different research fields where solving eigenvalue problems is the main technical task.

Acknowledgement

SDC would like to express his great gratitude to Profs. Kono and Fujimura for their generous hospitality and many exciting discussions during his short stay at Tohoku University, where the KEP idea was first formed. The authors thank National Taiwan University for financial support through the CQSE 10R80914-1 and National Science Council through NSC 100-2113-M-002-006-MY3.

**Reference**

[1] L. D. Landau and E. M. Lifschitz, *Quantum Mechanics* (Butterworth–Heinemann, Oxford, 1977).

[2] F. Rellich, *Perturbation Theory of Eigenvalue Problems* (Gordon and Breach, New York, 1969).

[3]O. Steinmann, *Perturbation Expansions in Axiomatic Field Theory* (Springer-Verlag, Berlin, 1971).

[4] F. M. Fernandez, Introduction to Perturbation Theory in Quantum Mechanics (CRC, Boca Raton, 2000).

[5] J. Cizek and E. R. Vrscay, *Int. J. Quan. Chem.* **21**, 27 (1982).

[6] A. R. P. Rau, *Astronomy-Inspired Atomic and Molecular Physics* (Kluwer Academic, Dordecht, 2002).

[7] A. Dalgarno and G. W. F. Drake, *Chem. Phys. Lett.* **3**, 349 (1969).

[8] H. R. Reiss, *Phys. Rev. A* **22**, 1786 (1980).

[9] V. G. Gueorguiev, W. E. Ormand, C. W. Johnson, and J. P. Draayer, *Phys. Rev. C* **65**, 024314 (2002).

[10] H.C. Bowen and J.W. Linnett, *Mol. Phys.* **6**, 387 (1963).

[11] Gaussian 09, Revision A.1, M.J. Frisch *et al*, Gaussian, Inc., Wallingford CT, 2009.

[12] I. N. Levine, *Quantum Chemistry* (5th ed.) Prentice-Hall (2000).

[13] D.R. Bates, K. Ledsham, A.L. Stewart, *Phil. Trans. Roy. Soc. A* **246**, 215 (1953).

[14] L. Pauling, *Chem. Phys.* **5**, 173 (1928).

[15] B.N. Finkelstein and G.E. Horowitz, *Z. Phys.* **48**, 118 (1928).

[16] M.M. Madsen and J.M.Peek, *Atom. Data* **2**, 171 (1971).

[17] L.J. Schaad and W.V. Hicks, *J. Chem. Phys.* **53**, 851 (1970).

Appendix I: Perturbative approach

In the case $V_1(x) \approx V_2(x)$, we can use perturbative approach. Consider first the two-state approximation.

$$C_1(2E_1 - E) + C_1\langle\psi_1|(V_2 - V_1)|\psi_1\rangle + C_2(E_2 + E_1 - E)\langle\psi_1|\psi_2\rangle = 0 \tag{37a}$$

$$C_2(2E_2 - E) + C_2\langle\psi_2|(V_1 - V_2)|\psi_2\rangle + C_1(E_1 + E_2 - E)\langle\psi_2|\psi_1\rangle = 0 \tag{37b}$$

The exact solution can be written as

$$E = E_1 + E_2 + \frac{f_1 + f_2}{2\left(1-\eta^2\right)} \pm \frac{\sqrt{4\left(1-\eta^2\right)\left[\left(E_1 - E_2\right)^2 + \left(E_1 - E_2\right)\left(f_1 - f_2\right) - f_1 f_2\right] + \left(f_1 + f_2\right)^2}}{2\left(1-\eta^2\right)} \tag{38}$$

with $f_1 = \langle\psi_1|(V_2 - V_1)|\psi_1\rangle$ and $f_2 = \langle\psi_2|(V_1 - V_2)|\psi_2\rangle$. Here only the lower bound state solution is adopted. Following the usual perturbative expansions; i.e.,

$$\begin{aligned}
\psi_2 &= \psi_1 + \lambda\psi_2^{(1)} + \cdots, \\
V_2 &= V_1 + \lambda V_2^{(1)} + \cdots, \\
E_2 &= E_1 + \lambda E_2^{(1)} + \cdots, \\
E &= E^{(0)} + \lambda E^{(1)} + \cdots
\end{aligned}$$

and $C_i = C_i^{(0)} + \lambda C_i^{(1)} + \cdots$ ($i$=1, 2), we obtain

$$\begin{aligned}
&\left(C_1^{(0)} + \lambda C_1^{(1)}\right)\left(2E_1 - E^{(0)} - \lambda E^{(1)} + \lambda\langle\psi_1|V_2^{(1)}|\psi_1\rangle\right) \\
&+\left(C_2^{(0)} + \lambda C_2^{(1)}\right)\left(2E_1 + \lambda E_2^{(1)} - E^{(0)} - \lambda E^{(1)}\right)\left(1 + \lambda\langle\psi_1|\psi_2^{(1)}\rangle\right) = 0
\end{aligned} \tag{39a}$$

$$\begin{aligned}
&\left(C_2^{(0)} + \lambda C_2^{(1)}\right)\left(2E_1 + 2\lambda E_2^{(1)} - E^{(0)} - \lambda E^{(1)} + \lambda\langle\psi_2|-V_2^{(1)}|\psi_2\rangle\right) \\
&+\left(C_1^{(0)} + \lambda C_1^{(1)}\right)\left(2E_1 + \lambda E_2^{(1)} - E^{(0)} - \lambda E^{(1)}\right)\left(1 + \lambda\langle\psi_2^{(1)}|\psi_1\rangle\right) = 0
\end{aligned} \tag{39b}$$

Up to the first order the solutions are $E^{(0)} = 2E_1$ and $E^{(1)} = f_1$. Notice that this result is the same as the energy calculated up to the 1st order in ordinary perturbation theory by using $H = H_0 + H'$ with $H_0 = T + 2V_1 = 2(T_1 + V_1)$ and $H' = V_2 - V_1$. Therefore, formally we recover the usual perturbation theory self-consistently. As a numerical demonstration, consider the two harmonic potentials case, $V_1(x) = \frac{kx^2}{2}$ and $V_2(x) = \frac{(k+\Delta k)(x+\Delta x)^2}{2}$. The solutions for both the subsystems and the whole system are well known. Using the dimensionless parameters, $M^* = 2M = 2,\ k = 1$, the distorted $(\Delta k = 0.1,\ \Delta x = 0)$ and the displaced $(\Delta k = 0,\ \Delta x = 0.1)$ oscillators are considered separately. For the distorted oscillator, we obtain the exact energy $E$=0.7246 , the KEP energy $E^{\mathrm{KEP}}$=0.7246, the zeroth order energy $E^{(0)}$=0.7070, and the perturbation energy $E^{(0)}+E^{(1)}$=0.7248. Similarly, for the displaced oscillator, we obtain $E$=0.7096 , $E^{\mathrm{KEP}}$=0.7096, $E^{(0)}$=0.7070, and $E^{(0)}+E^{(1)}$=0.7121. Therefore, we demonstrate that the perturbative approach does work for this case.

Appendix II: The time-dependent KEP formulation

To solve the time dependent Schroedinger equation

$$i\hbar \frac{\partial \Phi}{\partial t} = \hat{H}\Phi \tag{40}$$

we use the KEP scheme

$$\hat{H} = \hat{H}_1 + \hat{H}_2 \tag{41}$$

where

$$\hat{H}_1 = \frac{\hat{P}^2}{2(2m)} + V_1 \tag{42a}$$

$$\hat{H}_2 = \frac{\hat{P}^2}{2(2m)} + V_2 \tag{42b}$$

Assume the subsystem solutions are known

$$i\hbar \frac{\partial \psi_{1n}}{\partial t} = \hat{H}_1 \psi_{1n} \tag{43a}$$

$$i\hbar \frac{\partial \psi_{2k}}{\partial t} = \hat{H}_2 \psi_{2k} \tag{43b}$$

The total wavefunction can be expanded as

$$\Phi = \sum_n C_{1n}\psi_{1n} + \sum_k C_{2k}\psi_{2k} \tag{44}$$

where the expansion coefficients $C_{1n}$ and $C_{2k}$ depend on time. Substituting this into the Schroedinger equation and using the usual multiplication-integration procedure, we have

$$i\hbar \frac{dC_{1m}}{dt} + \sum_k i\hbar \frac{dC_{2k}}{dt} \langle \psi_{1m} | \psi_{2k} \rangle = \sum_k C_{2k} \langle \psi_{1m} | \hat{H}_1 | \psi_{2k} \rangle + \sum_n C_{1n} \langle \psi_{1m} | \hat{H}_2 | \psi_{1n} \rangle \tag{45a}$$

$$i\hbar \frac{dC_{2\ell}}{dt} + \sum_n i\hbar \frac{dC_{1n}}{dt} \langle \psi_{2\ell} | \psi_{1n} \rangle = \sum_n C_{1n} \langle \psi_{2\ell} | \hat{H}_2 | \psi_{1n} \rangle + \sum_k C_{2k} \langle \psi_{2\ell} | \hat{H}_1 | \psi_{2k} \rangle \tag{45b}$$

These are the working equations for the time dependent KEP method.

Figure captions

Fig. 1. Comparison of $K^{KEP}$ using the KEP solution scheme and the exact K as a function of the well depth $\lambda$. Here we use the dimensionless $m=a=1$.

Fig. 2. Comparison of the KEP and the exact wavefunctions for $\lambda$=0.2 using the dimensionless units.

Fig. 3. Comparison of the calculated energy levels. The first two columns from the left represent the energy levels determined from the subsystem Hamiltonian $\hat{H}_1$ and $\hat{H}_2$, respectively. The third (pink bars) represents the exact solutions. The others are the KEP solutions with the number of states from $N$=1 to $N$=5. Some higher energy levels are too far from the exact solutions, especially for small $N$, to be shown for better visibility.

Fig. 4. The total wavefunctions at (a) the ground state n=0; $\Psi_{n=0}(x)$, (b) the first excited state n=1; $\Psi_{n=1}(x)$, (c) the second excited state n=2; $\Psi_{n=2}(x)$ and (d) the third excited state n=3; $\Psi_{n=3}(x)$. The black line corresponds to the exact wave

function and the others correspond to the wave functions calculated with the number of states $N$=1, 3, and 5.

Fig. 5. The $H_2^+$ ground state potential energy curves calculated by Bowen et al. [10], Gaussian09 program package [11], the LCAO-MO method [12], and the KEP method. The parameter $\alpha$ as a function of R is shown in the insert.

Fig. 6. The $H_2^+$ ground state wavefunctions calculated by Bowen et al. [10], Gaussian09 program package [11], and the KEP method at the nuclear distance $R$=1.0 a.u. and $R$=2.0 a.u. where $y$=$z$=0.

Table caption

Table 1. Comparison of the equilibrium bond distances and bond energies calculated by the LCAO-MO [12], Bates *et al* [13], Pauling [14], Bowen *et al* [10], Finkelstein *et al* [15], Madsen *et al* [16], Gaussian09 program package [11], KEP and experimental data [17].

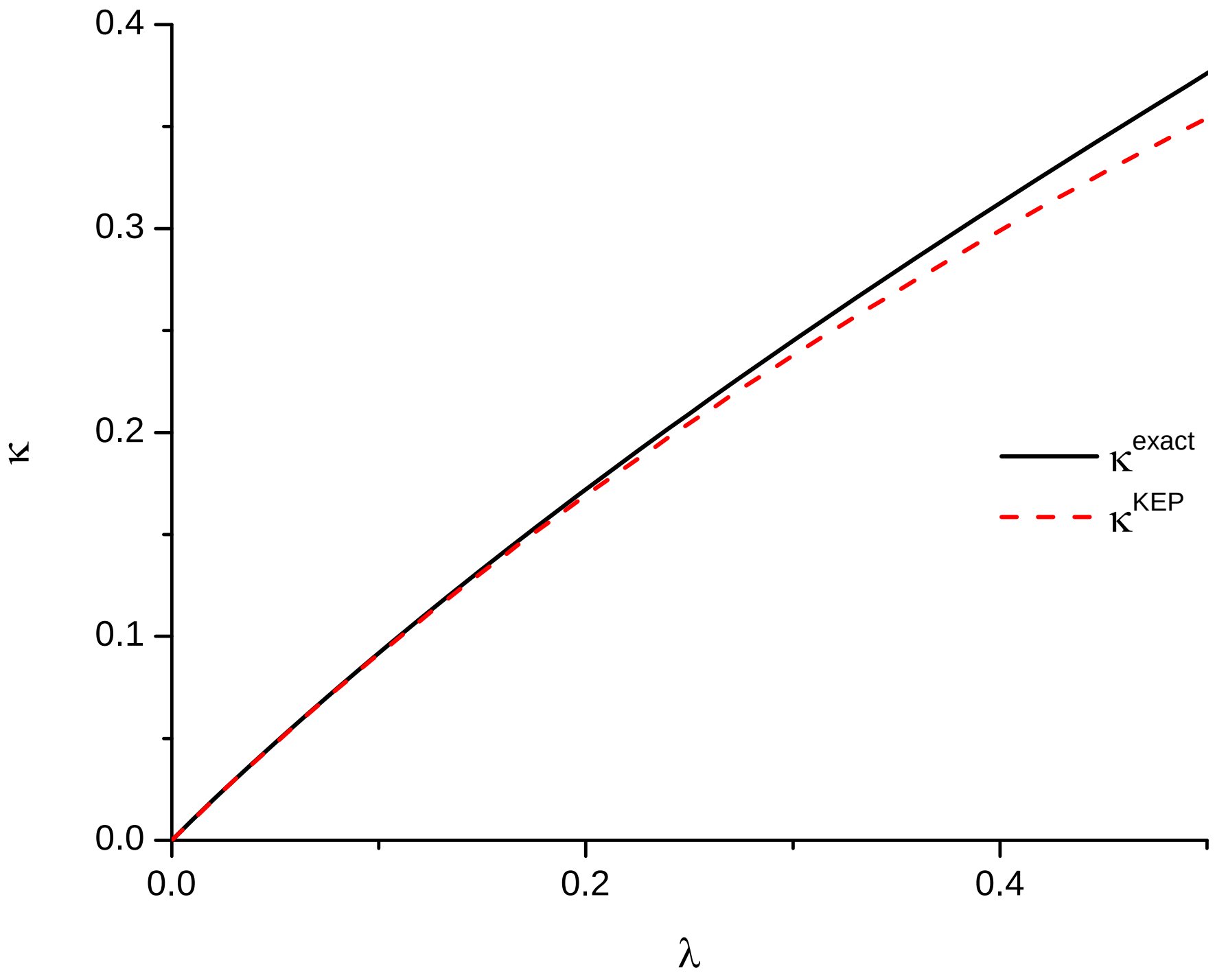

Fig. 1

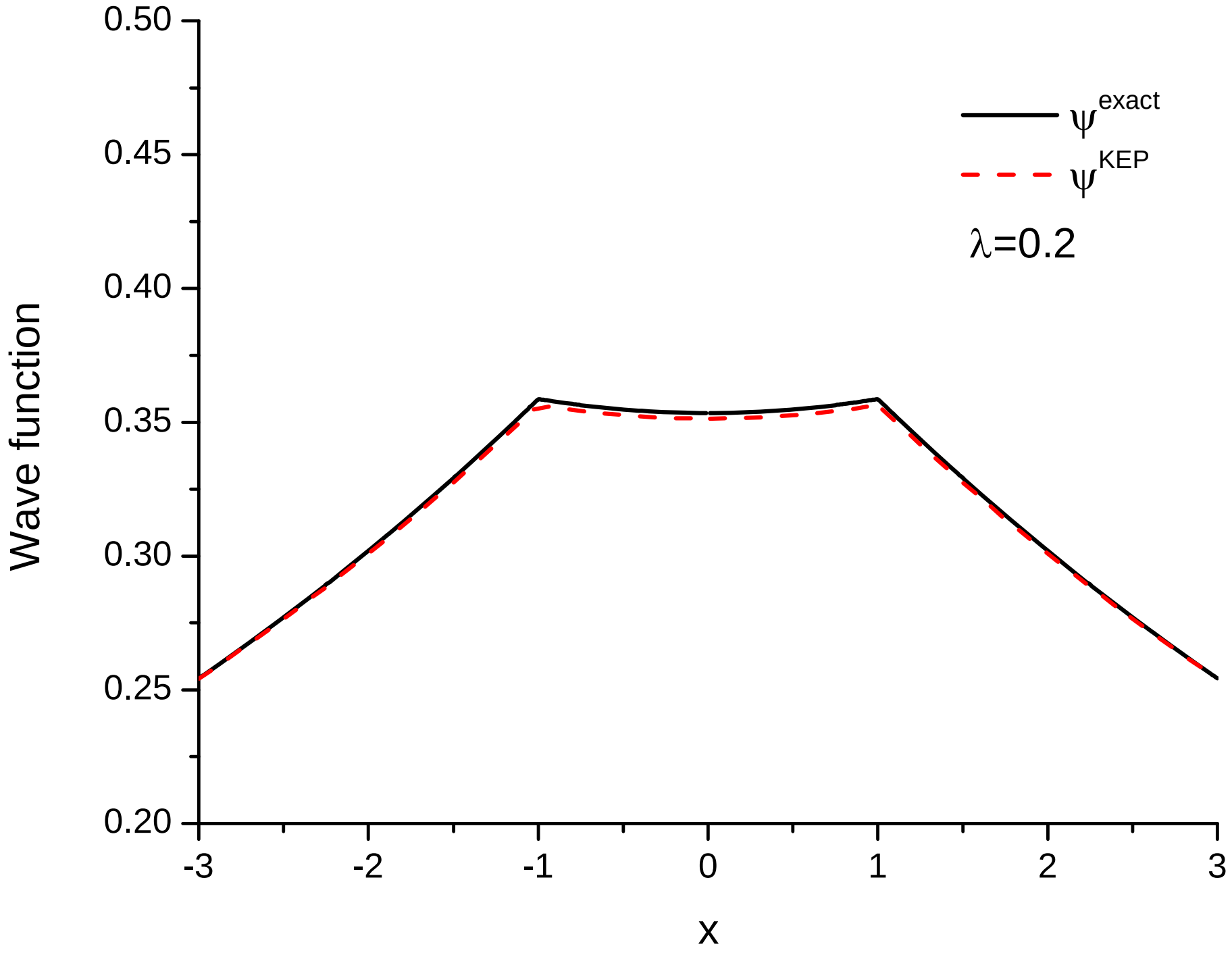

Fig. 2

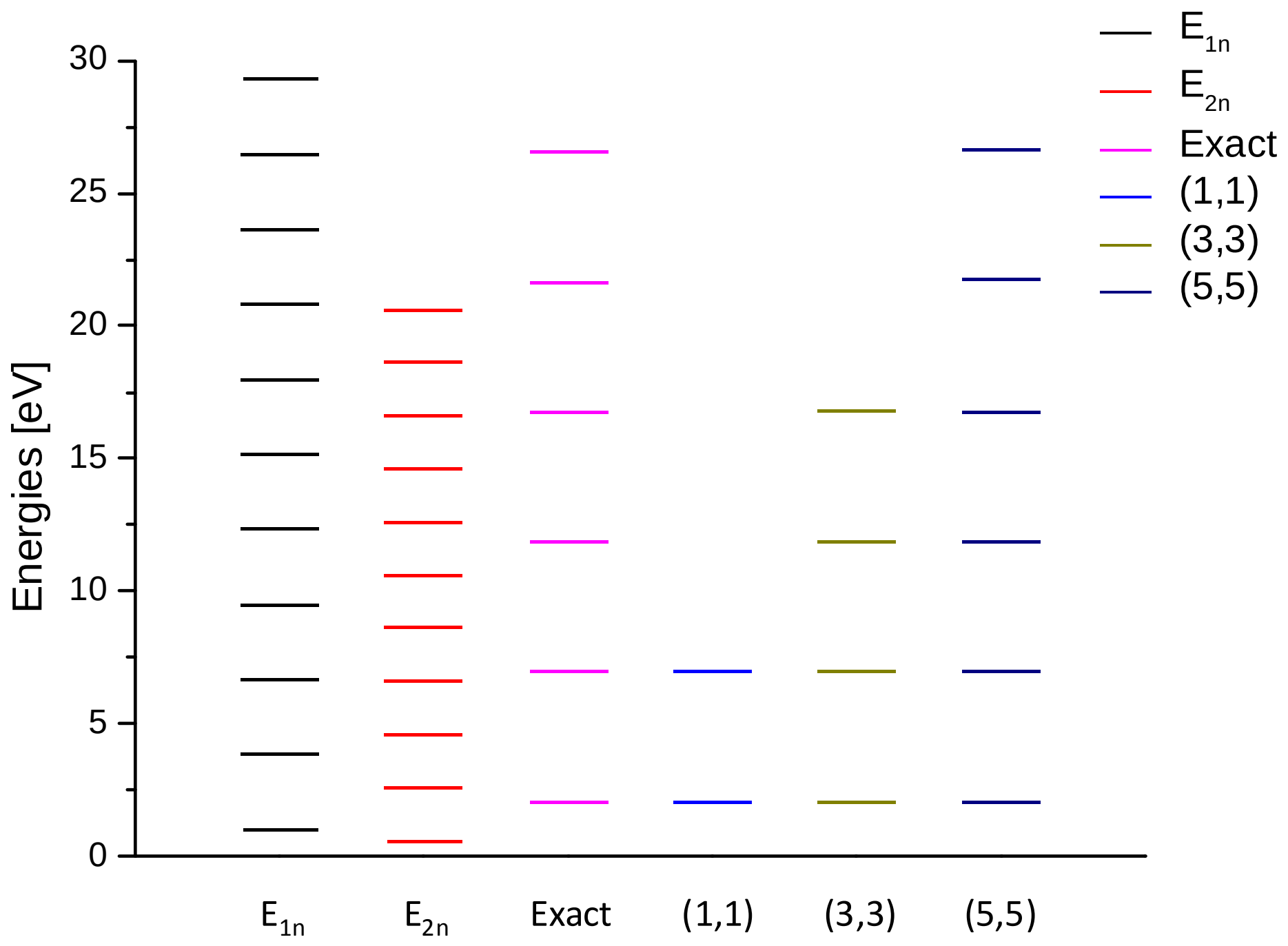

Fig. 3

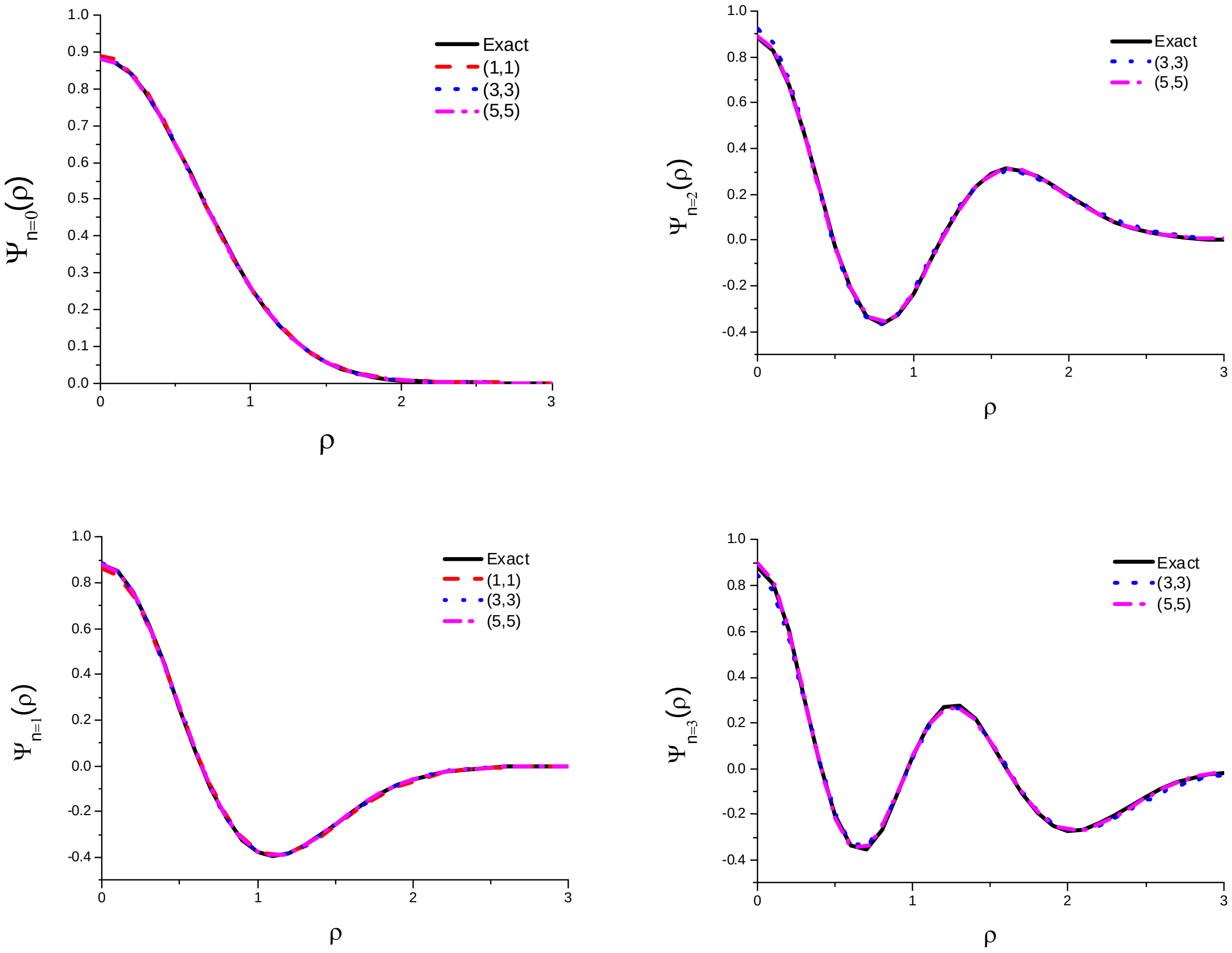

Fig. 4

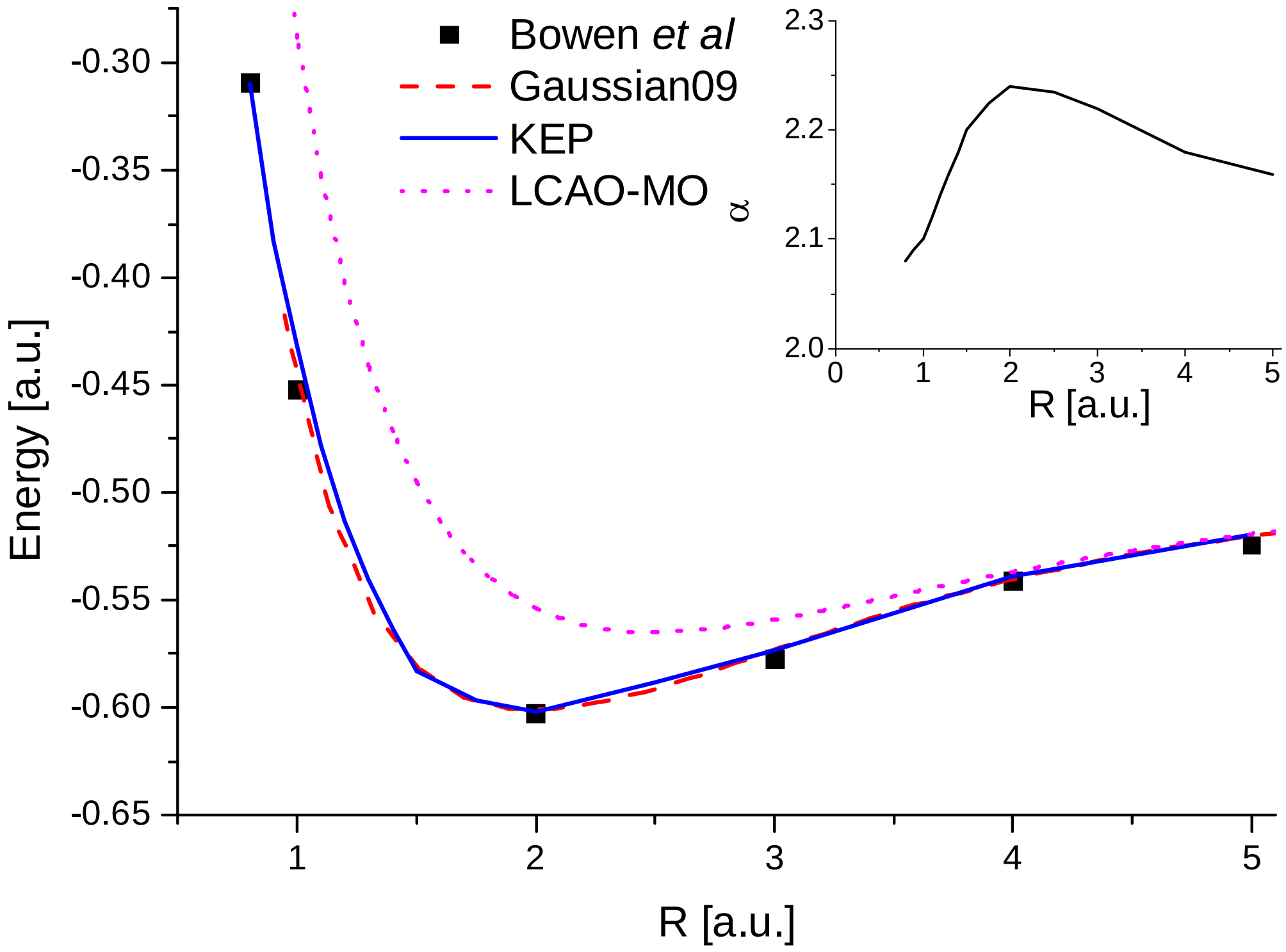

Fig. 5

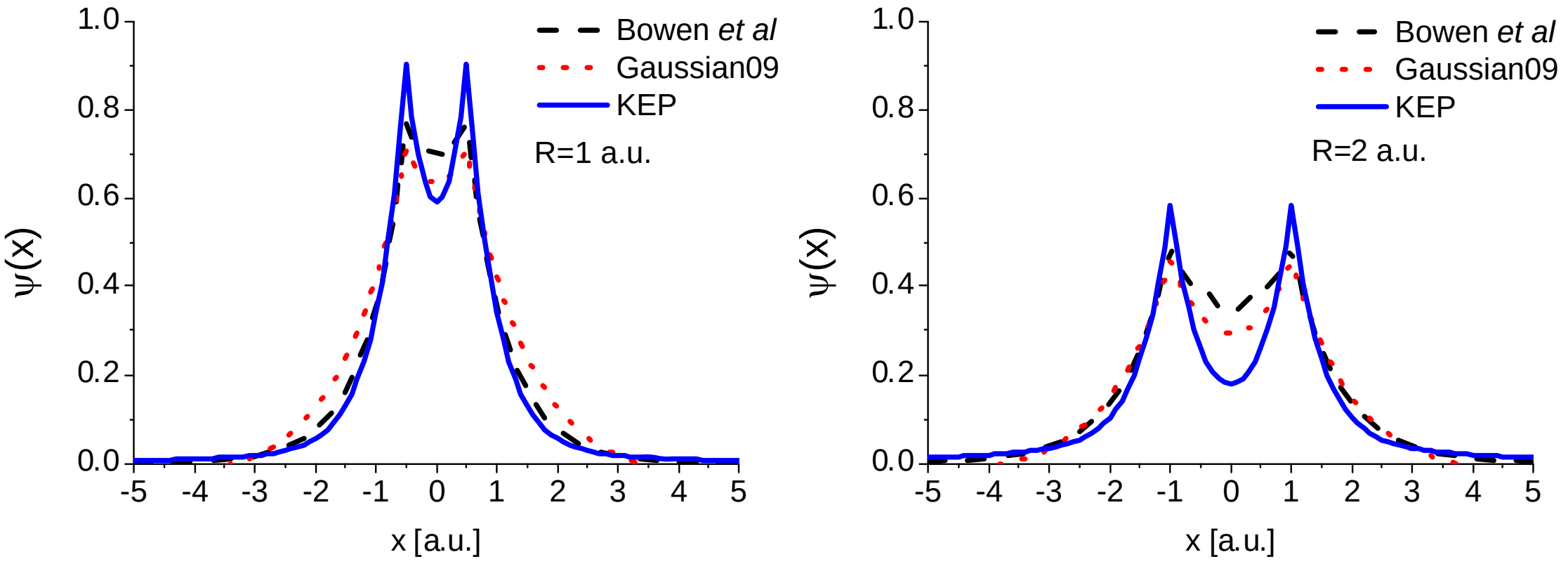

Fig. 6

| | $R_e$ [a.u.] | $E_e$ [a.u.] |
|---|---|---|
| LCAO-MO [12] | 2.5 | -0.5648 |
| Bates *et al.* [13] | 2.0 | -0.6026 |
| Pauling [14] | 2.5 | -0.5648 |
| Bowen *et al* [10] | 2.0 | -0.5985 |
| Finkelstein *et al* [15] | 2.0 | -0.5865 |
| Madsen *et al* [16] | 2.0 | -0.6026 |
| Gaussian09 [11] | 2.0 | -0.6012 |
| KEP | 2.0 | -0.6019 |
| Experiment [17] | 2.0 | -0.6026 |

Table 1